\documentclass[aps,pre,twocolumn,showpacs,eqsecnum]{revtex4}
\usepackage{amssymb}
\usepackage{amsmath}

\newcommand{\bq}{\begin{equation}}
\newcommand{\nq}{\end{equation}}

\begin{document}

\title{Non-Markovian Stochastic Liouville equation and its Markovian
representation. Extensions of the continuous time random walk
approach.}
\author{A. I. Shushin}
\affiliation{Institute of Chemical Physics, Russian Academy of
Sciences, 117977, GSP-1, Kosygin str. 4, Moscow, Russia}

\begin{abstract}
Some specific features and extensions of the continuous time random
walk (CTRW) approach are analyzed in detail within the Markovian
representation (MR) and CTRW-based non-Markovian stochastic
Liouville equation (SLE). In the MR CTRW processes are represented
by multidimensional Markovian ones. In this representation the
probability distribution function (PDF) $W (t)$ of fluctuation
renewals is associated with that of reoccurrences in a certain jump
state of some Markovian controlling process. Within the MR the
non-Markovian SLE, which describes the effect of CTRW-like noise on
relaxation of dynamic and stochastic systems, is generalized to take
into account the influence of relaxing systems on statistical
properties of noise. The generalized non-Markovian SLE is applied to
study two modifications of the CTRW approach. One of them considers
the cascaded CTRWs in which the controlling process is actually
CTRW-like one controlled by another CTRW process, controlled in turn
by the third one, etc. Within the MR simple expression for the PDF
$W (t)$ of total controlling process is obtained in terms of
Markovian variants of controlling PDFs in the cascade. The
expression is shown to be especially simple and instructive in the
case of anomalous processes determined by long time tailed $W (t)$.
The cascaded CTRWs can model the effect of complexity of a system on
relaxation kinetics (in glasses, fractals, branching media,
ultrametric structures, etc.). Another CTRW-modification describes
the kinetics of processes governed by fluctuating $W (t)$. Within
the MR the problem is analyzed in a general form without restrictive
assumptions on correlations of PDFs of consecutive renewals. The
analysis shows that $W (t)$ can strongly affect the kinetics of the
process. Possible manifestations of this effect are discussed.
\end{abstract}

 \maketitle

\bigskip

\section{Introduction}

Relaxation and transport phenomena in condensed media are of great
importance for understanding of various processes in physics,
chemistry, biology etc. \cite{Gard,Fors,Abr}. For description of
these phenomena a large number of approaches  have been proposed in
literature \cite{Gard,Fors,Hau,Bou}. The most popular are Markovian
approaches based on Langevin and Fokker-Plank \cite{Gard,Fors}
equations assuming negligibly short memory in processes under study.
Great advantage of these approaches consists in possibility of
relatively simple treatment of relaxation in dynamical and
(Markovian) stochastic systems induced by interactions whose
fluctuations are Markovian processes (MPs). This type of relaxation,
which in what follows for brevity is called MP affected MPs, is
known to be described by the stochastic Liouville equation (SLE)
\cite{Kubo}.

Recent investigations show, however, that Markovian theories do not
properly describe the kinetics of relaxations and fluctuations of
various observables in many non-equilibrium processes
\cite{Bou,Met}. The non-Markovian relaxation kinetics in these
processes results form strong memory effects. Moreover, in a number
of systems the kinetics is anomalously long time tailed which
implies long time memory in them. The anomaly manifests itself in
some interesting phenomena such as non-ergodicity
\cite{Met,Bark0,Bark1}, aging \cite{Grig00,Grig01,Bark2}, slow
relaxation \cite{Met,West,Grig02,Grig1}, etc.

Anomalous specific features of relaxation in highly non-equilibrium
systems attract close attentions of increasing number of scientists
\cite{Met,West,Grig1}. Analysis of these features can be made within
different models and approximations. One of the most popular and,
probably, adequate is continuous time random walk (CTRW) approach in
which the process is represented by jump-like fluctuations (or
relaxation) treated as a sequence of renewals \cite{Scher1,Met,Hau}
characterized by the probability density function (PDF) $W(t)$ of
waiting times of renewals [this PDF is often denoted as $\psi (t)$
\cite{Wei,Scher1,Met}].

Last years different variants have been analyzed and applied to
treating a large number of processes \cite{Met,West}. In some recent
studies special attention have been paid to above-mentioned
anomalously long time tailed relaxation \cite{Met,Grig1}: within the
CTRW approach such behavior is assumed to result from that of the
PDF $W(t) \sim 1/t^{1+\alpha}$ with $\alpha < 1$ \cite{Met,West1}.
The applications of the CTRW approach allowed one to interpret a lot
of experimental results \cite{Met,Grig1}. It is shown, in
particular, that in many cases the specific properties of
fluctuations and relaxation can quite properly be described within
the anomalous CTRW approach (with anomalous behavior of the PDF
$W(t)$).

The CTRW approach is fairly popular last years. It is successfully
applied both to general investigations and to interpretation of
experimental results. It is worth pointing out, however, that in
majority of most interesting applications the validity of this
approach is not quite evident. Typically, it is justified with the
use of intuitive arguments concerning statistical properties of
systems under study. In addition, some recent investigations show
that, despite great flexibility, the conventional CTRW approach
proves to be not quite accurate in description of the kinetics of a
number of non-Markovian processes \cite{Klaf1,Bark2}. In such a case
any further extensions of this approach are of course desirable.

Certain modifications of the conventional CTRW approach have already
been discussed in literature. In some CTRW-variants the
non-homogeneity of the process, consisting in the dependence of the
jump-like fluctuations on the fluctuation number, have been taken
into account \cite{Shu1,Sok1}. In other variants the modification of
simple CTRW time sequences of renewals is proposed \cite{Grig2}.
These modifications are very interesting and essentially clarify
specific features of CTRW-like processes.

In this work we will propose and analyze some extensions of the CTRW
approach within the Markovian representation \cite{Shu1,Shu2,Shu3}.
This method is based on idea that under fairly general assumptions a
wide variety of non-Markovian CTRW processes can be treated as MPs
with fluctuating parameters, whose fluctuations are described by
other MPs called hereafter controlling processes. In so doing these
CTRW processes are unambiguously represented by some
multidimensional MPs \cite{Shu1} and the statistics of renewals is
associated with that of reoccurrences in certain transition state
(or the state of onset of fluctuation jumps) during the controlling
MPs in some auxiliary spaces. In particular, the PDF $W(t)$ is
expressed in terms of the characteristic functions of the
controlling processes.

The Markovian representation is shown to be very useful for the
analysis of non-Markovian CTRW-like models describing different
physical processes \cite{Shu1,Shu2,Shu3}. In addition it permits
rigorous derivation of the non-Markovian stochastic Liouville
equation (SLE) treating relaxation in dynamical and Markovian
stochastic systems induced by CTRW-type fluctuating interactions
with thermal bath \cite{Shu1,Shu2,Shu3}. In what follows, for
brevity, Markovian systems in which relaxation is induced by
CTRW-fluctuating interactions will be called CTRW affected MPs.

In this work the compact formulation of the Markovian representation
is proposed which simplifies and generalizes derivation of the
CTRW-based non-Markovian SLE thus allowing for extension of the SLE
to take into account back effect of fluctuating system on effective
statistical properties of fluctuations. This formulation especially
clearly demonstrates that CTRW processes can be considered as MP
affected MPs (in above-proposed brief terminology) and basic
equations of the CTRW approach are nothing else but the SLEs in a
reduced form.

The obtained generalized non-Markovian SLE is applied to the
analysis of validity of CTRW approaches. Two important extensions of
the CTRW approach are proposed and discussed in detail:

The first modification describes the effect of cascaded controlling
processes in which the process controlling renewals is assumed to be
CTRW-like one controlled by the second CTRW process, which in turn
is controlled by the third CTRW processes, etc. In this cascaded
CTRW control model the compact representation for the PDF $W(t)$
(more correctly for the Laplace transform of this function) is
obtained in terms of PDFs of controlling processes $W_j(t)$ at all
cascade steps $j$, found assuming these processes to be Markovian
(i.e. neglecting control). This model is very suitable for the
analysis of non-Markovian relaxation kinetics in structured and
disordered systems especially in the case of anomalous long time
tailed behavior of PDFs $W_j(t)$.

The second modification treats CTRW-like processes governed by
fluctuating PDFs $W(t)$. This type of processes can be considered as
an extension of conventional CTRW-approach in which fluctuations of
$W(t)$ are assumed to result from the additional effect of
non-equilibrium medium and are modeled by the dependence of the
system on the Markovian stochastic variable of special type. In this
model and within the Markovian representation the description of
these processes reduces to solving the non-Markovian SLE. The
analysis of the model shows that fluctuations of the PDF $W(t)$ can
strongly change the kinetics of CTRW relaxation. Especially
significant effect is expected in the case of anomalous long time
tailed processes.

\section{General formulation}

We consider relaxation processes in a dynamical or stochastic
Markovian system induced by fluctuating interaction with the
classical thermal bath. The Markovian evolution of the system is
assumed to be governed by the fluctuating Liouville operator ${L}
(t)$. Our general formulation is applicable both to quantum and
classical systems. In particular, in the dynamic systems the only
difference of these two cases is in the form of the operators ${L}$:
for a quantum system ${ L} = i[H,\dots]$, (here $[H,\dots]$ is the
commutator with the Hamiltonian $H$ of the system), while for a
classical system ${L} = \{H,\dots\}$ (with $\{H,\dots\}$ being the
classical Poisson brackets). In what follows, for definiteness and
brevity, we will concentrate on general results as applied to
classical systems.

Fluctuations of $L (t)$ are assumed to result from the dependence on
the fluctuating bath coordinate ${\bf x}$: $L (t) \equiv L_{{\bf
x}(t)}$, whose changes are modeled by stochastic jumps between
states $|\nu \rangle \equiv |{\bf x}_{\nu} \rangle$ with different
$L = L_{\nu}$ in the multidimensional space $ \{ x \}$.

Hereafter we will use "bra-ket" notation for the states in $\{ x
\}$-space  suitable for treating relaxation phenomena determined by
not self-adjoint evolution operators.

In the model under study the system evolution is described by the
(PDF) $\rho (t)$ (or density matrix for quantum systems) which
satisfies the linear equation
\begin{equation}\label{genf1}
\dot \rho = - \hat L \rho \;\;\mbox{with}\;\; \hat L \equiv  \hat
L_{\bf x} = \sum\nolimits_{\nu} \! |{\bf x}_{\nu} \rangle L_{\nu}
\langle {\bf x}_{\nu}| .
\end{equation}
For simplicity of presentation, $\hat L$ is considered to be
diagonal in $\{ x \}$-space though the formulas obtained are valid
in the case non-diagonal $\hat L$ as well.

The operator solution of eq. (\ref{genf1}) is represented as
\begin{equation}\label{genf2}
\rho(t) = \hat U(t)\rho_0 \;\, \mbox{with} \,\; \hat U(t) = \hat T
\Big[ e^{-\!\int_0^{t}\!d\tau \hat {L}(\!\tau\!)} \Big],
\end{equation}
where $\hat T$ is the time ordering operator and $\rho_0 = \rho (t =
0)$ is the initial condition.

Experimentally measured observables are usually described by the
evolution operator $\hat U (t)$ averaged over stochastic
fluctuations of ${\bf x}(t)$ which is expressed in terms of the
conditional evolution operator $\hat {\mathbb G}(x,x'|t)$ as
\begin{equation}
\langle \hat U(t) \rangle_{\bf x} = \sum\nolimits_{{\bf x},{\bf
x}_0} \!\! \hat{\mathbb G} ({\bf x},{\bf x}_0|t)\sigma_i ({\bf x}_0)
\equiv \langle {\bf x}_e| \hat{\mathbb G}|{\bf x}_i\rangle ,
\label{genf3}
\end{equation}
where $\sigma_i ({\bf x}) \equiv |{\bf x}_i\rangle $ is the initial
PDF of the system in $\{x\}$-space normalized by the condition
$\sum_{\bf x} \sigma_i ({\bf x}) = 1$. In eq. (\ref{genf3}) we have
also introduced the (adjoined) equilibrium state vector which in
bra-ket notation is represented as [see below eq. (\ref{genf4a})]:
$\langle {\bf x}_e| = \sum\nolimits_{\nu}\langle {\bf x}_{\nu}|$.

In accordance with eq. (\ref{genf3}) the problem reduces to
evaluating the operator $\hat {\mathbb G} ({\bf x},{\bf x}_i|t)$.
Unfortunately this can be done relatively easily only for very few
models of stochastic ${\bf x}(t)$-fluctuations. In this work we will
analyze some of those models based on the CTRW-approach.

\section{Models of fluctuations}

\subsection{Markovian models}

The Markovian approach is based on the assumption that $\hat
L(t)$-fluctuations are described by the stochastic MPs in
$\{x\}$-space and  $\hat L(t)$-evolution is described by the PDF
$\sigma ({\bf x},{\bf x}_0|t)$ satisfying equation
\begin{equation}
\dot \sigma = - \hat {\cal L} \sigma \;\;\; \mbox{with} \;\;\;
\sigma ({\bf x},{\bf x}_0|0) = \delta_{{\bf x},{\bf x}_0},
\label{mark1_0}
\end{equation}
where $\hat {\cal L}\equiv \hat {\cal L}_{{\bf x}} $ is some linear
operator in $\{x\}$-space. The principal simplification of the
problem results from the fact that in the Markovian approach
(\ref{mark1_0}), i.e. in the case of MP affected MPs, $\hat
{\mathbb{G}} ({\bf x},{\bf x}_0|t)$ obeys the SLE:
\begin{equation}
\dot {\hat {\mathbb{G}}} = -(\hat {L} + \hat {\cal L}) {\hat
{\mathbb{G }}}, \;\;\:\,\mbox{so that} \;\;\,{\hat {\mathbb{G}}}(t)
= e^{-(\hat {L} + \hat {\cal L})t}. \label{mark2_0}
\end{equation}
which yields for $\langle\hat {\widetilde {U}}\rangle =
\int_0^{\infty} \! dt \, \langle\hat U(t)\rangle \exp (-\epsilon
t):$
\begin{equation}
\langle\hat {\widetilde {U}}\rangle  = \langle \hat {\widetilde
{\mathbb{G}}} \rangle_{\bf x} = \langle {\bf x}_e| (\epsilon + \hat
L + \hat {\cal L})^{-1}|{\bf x}_i\rangle.
 \label{mark2_0a}
\end{equation}

In the form  (\ref{mark2_0}) the SLE  is valid for any dependence
$\hat {L }_{\bf x}$ on the coordinate ${\bf x}$ though, in general,
it is still very complicated for analysis either numerical or
analytical.

Significant simplification can be gained within some special models,
for example, in the CTRW approach.

\subsection{CTRW-based models}

Non-Markovian $\hat L (t)$-fluctuations can conveniently be
described by the CTRW approach \cite{Hau,Met,Wei}. It treats
fluctuations as a sequence of sudden, jump-like changes of $\hat L
$. In the simplest variant of the CTRW (more complicated variants
are discussed below) the onset of any particular change of number
$\beta$ is described by the probability $P_{\beta -1} (t)$ (in
$\{x\}$-space) not to have any change during time $t$ and its
derivative $ W_{\beta -1} (t) =-\dot {P}_{\beta -1} (t)$, i.e. the
PDF for times of waiting for the change. These functions are
independent of $\beta$ for $\beta > 1$ and for $\beta = 1$ depend on
the problem considered \cite{Hau,Met}:
\begin{equation}
W_{\beta \geq 1} (t) = W_n (t), \;\; W_{0} (t) = W_i (t) ,
\label{nmark1}
\end{equation}
and ${ P}_{\!j} (t) = \int_0^t d \tau \, W_j (\tau), \, (j = n,i),$.
In what follows we will mainly discuss the non-stationary CTRW
variant in which $W_{\beta \geq 0} (t) = W_n (t)$.

The Laplace transforms of $W_j (t) $ and $P_j (t), \, (j = n,i)),$
are related by simple equation $ { \widetilde P}_{\!j} (\epsilon)=
[1 - {\widetilde W_{\!j}} (\epsilon)]/\epsilon$ with
\begin{equation}
{\widetilde W}_{\!j} (\epsilon) = [1 + \Phi_{\!j} (\epsilon)]^{-1} ,
\; (j = n,i).\label{nmark4}
\end{equation}
In eq. (\ref{nmark4}) $\Phi_{\!j} (\epsilon)$ is the important
auxiliary function \cite{Shu1}.

\subsubsection{Markovian representation}

The results of our earlier studies \cite{Shu1,Shu2}  show that
important non-Markovian generalizations of the SLE (\ref{mark2_0})
can be obtained by assuming the operator $\hat {\cal L} (t)$ to be a
stochastic function of time.

In this work we will analyze quite natural generalization of the
Markovian SLE (\ref{mark2_0}) based on the assumption that
fluctuations of $\hat {\cal L} (t)$ are Markovian. In other words
$\hat {\cal L} (t)\equiv \hat {\cal L}_{{\bf z}(t)}$ is assumed to
be a function of the Markovian stochastic (controlling) variable
${\bf z} (t)$ which, in general, is a vector, so that generalizing
the solution (\ref{mark2_0}) of the Markovian SLE one can write
\begin{equation}
{\hat {\mathbb{G}}}(t) = \hat T \{\exp [-\mbox{$\int_0^t\!d\tau(\hat
{L}_{{\bf x}(\tau)} + \hat {\cal L}_{{\bf z}(\tau)})$}]\}.
\label{genf4_0}
\end{equation}

In the major part of the analysis we will assume that both $\{x\}$-
and $\{{z}\}$-spaces are discrete and use bra/ket notation with
Greek and Latin characters for states in these two spaces, for
example, $|{\bf x}_{\nu} \rangle$ and $|{\bf z}_{j}\rangle$,
respectively (although, in case of need the continuous variants of
$\{x\}$- and $\{{z}\}$-spaces will also be considered).

We start our analysis with the simple Markovian stochastic migration
process in $\{x\}$-space, which is described by the Kolmogov-Feller
jump matrix
\begin{equation}
\hat{\cal L} =  \hat k^{d}_{{\bf z}\!(t)}\!-\!\hat P_x\hat
k^{n}_{{\bf z}\!(t)}\,\; \mbox{with}\, \; \hat P_x =
\mbox{$\sum_{\nu\neq\mu}$} p_{\nu\mu}|{\bf x}_{\nu}\rangle \langle
{\bf x}_{\mu}| \label{genf4}
\end{equation}
in which $\hat k^d_{{\bf z}\!(t)} \equiv \hat k^d[{\bf z}(t)]$ and
$\hat k^n_{{\bf z}\!(t)} \equiv \hat k^n[{\bf z}(t)]$ are the
matrices (diagonal in $\{x\}$-space) of $z$-dependent, i.e.
fluctuating in time, jump rates and $p_{\nu\mu} $ are the
probabilities of jumps $\{x\}$-space normalized by the relation
$\sum_{\nu}\! p_{\nu \mu} = 1$. For simplicity, we assume that $\hat
P_x$ is independent of $z$, though most general results obtained in
this section are valid in the case of $z$-dependent $\hat P_x$ as
well (see below).

The matrix $\hat{\cal L}(t)$ describes relaxation in $\{x\}$-space
to the equilibrium state
\begin{equation}\label{genf4a}
|{\bf x}_e\rangle = \mbox{$\sum\nolimits_{\nu}$} p_{\nu}^{e}|{\bf
x}_{\nu}\rangle, \;\;\mbox{with}\;\; \langle {\bf x}_e | =
\mbox{$\sum\nolimits_{\nu}$}\langle {\bf x}_{\nu}|,
\end{equation}
for which $(1 - \hat P_x)\hat k|{\bf x}_e\rangle = 0$ and $\langle
{\bf x}_e |(1 - \hat P_x)\hat k = 0$. This state is assumed to be
independent of ${\bf z}$. Noteworthy is that even in the absence of
the equilibrium state $|{\bf x}_e\rangle $, for example when
$\hat{\cal L}$ describes diffusive migration in infinite space, the
adjoint vector $\langle {\bf x}_e |$ defined in eq. (\ref{genf4a})
still exists and satisfies the relation $\langle {\bf x}_e |(1 -
\hat P_x)\hat k = 0$ which means nothing else but the conservation
of population in the process of migration in $\{x\}$-space.

Within the considered Markovian approximation for ${\cal L}_{{\bf
z}(t)}$-fluctuations the evolution of the system in controlling
$\{z\}$-space is governed by the PDF $\varphi ({\bf z},{\bf z}_0|t)$
satisfying equation
\begin{equation}
\dot \varphi = - \hat {\Lambda} \varphi \;\;\; \mbox{with} \;\;\;
\varphi ({\bf z}|0) = |\varphi_i({\bf z})\rangle \equiv |{\bf
z}_i\rangle, \label{mark1_1}
\end{equation}
in which $\hat {\Lambda}\equiv \hat \Lambda_{{\bf z}} $ is some
linear operator describing relaxation in $\{z\}$-space and
$\varphi_i({\bf z}) \equiv |{\bf z}_i\rangle$ is the initial state
of the controlling process (in this $\{z\}$-space) in the par-ket
representation. Similar to the operator ${\cal L}$ in $\{x\}$-space
for our further applications it is convenient to introduce the
equilibrium states of the operator $\hat {\Lambda}$
\begin{equation}\label{mark1_2}
|{\bf z}_e \rangle = \mbox{$\sum\nolimits_{j}$} p_{j}^{{\bf z}}|{\bf
z}_{j}\rangle \;\;\mbox{and}\;\; \langle {\bf z}_e | =
\mbox{$\sum\nolimits_{j}$}\langle {\bf z}_{j}|.
\end{equation}
As in the case of $\{x\}$-space considered above, the vector
$\langle {\bf z}_e |$ is still meaningful even in the absence of the
equilibrium state of the operator $\hat {\Lambda}$.

{\it The operator $\hat {\Lambda}$ can, in general, depend on ${\bf
x}$. Moreover,  $\hat {\Lambda}$ can be a non-diagonal matrix in $\{
x \}$-space though, for simplicity of further discussion, we will
assume $\hat {\Lambda}$ to be diagonal in this space}.

In our analysis, in case of need we will use the simple and flexible
continuous Smoluchowski model for controlling process in
$\{z\}$-space in which
\begin{equation}
\hat \Lambda =  - D_z
z^{1-n_z}e^{-u(z)}\nabla_z[z^{n_z-1}e^{u(z)}\nabla_z], \label{mod1}
\end{equation}
with $z = |{\bf z}|, \, u(z)$ is the effective potential, and
$\nabla_z = \partial/\partial z$, is the radial part of the operator
describing diffusion in $n_z$-dimensional $\{z \}$-space with the
diffusion coefficient $D_z$. This model is quite sufficient for our
qualitative and semiquantitative analysis. The correspondence
between continuous model implying smooth functions $D_z$ and $u(z)$
and its discrete variant considered above is formulated as follows:
${\bf z} \leftrightarrow b|{\bf z}_j\rangle$, where $b$ is the
spacing in the discrete model.

Similarly to the simplest model discussed in Sec. III.A one can
easily see that in this approximation the evolution operator $\hat
{\mathbb{G}}_{\bf r}({\bf r},{\bf r}_i|t)$, which determines
$\langle \hat {U} (t) \rangle$  obeys the Markovian SLE in the
extended space $\{{\bf r}\}$ including the dynamical subspace
$\{x\}$ and additional stochastic subspace $\{z\}$: $\{{\bf r}\} =
\{ x \otimes z \}$:
\begin{equation}
\dot {\hat {\mathbb{G}}}_{\bf r} = -(\hat { L} + \hat {\cal L} +
\hat {\Lambda}) \hat {\mathbb{G }}_{\bf r} \;\:\mbox{with} \;\;\hat
{\mathbb{G}}_{\bf r} ({\bf r},{\bf r}_i|0) = \delta_{{\bf r} {\bf
r}_i}, \label{mark2_1}
\end{equation}
so that we get for ${\hat {\mathbb{G}}}$ and the Laplace transform
$\hat {\widetilde {U}} = \langle \hat {\widetilde {\mathbb{G}}}
\rangle_x$:
\begin{equation}
{\hat {\mathbb{G}}} = \langle {\bf z}_e|{\hat {\mathbb{G}}}_{\bf
r}|{\bf z}_i\rangle \;\;\mbox{and}\;\; \langle \hat {\widetilde {U}}
\rangle = \langle (\hat \Omega + \hat {\cal L} + \hat
{\Lambda})^{-1}\rangle_{\bf r},
 \label{mark2_1a}
\end{equation}
where $\hat \Omega = \epsilon + \hat { L}$. In the form
(\ref{mark2_1a}) the SLE  is valid for any dependence of coupling
$\hat {\cal L }_{z}$ on the coordinate ${z}$ though, in general, it
is still very complicated for analysis either numerical or
analytical.

Significant simplification can be gained within some special models
for jump-rate matrix $\hat k_{z}$ [see eq. (\ref{genf3})]. One of
the most well known is the CTRW-model discussed below, in Sec. IVB.

\subsubsection{Localized jump rates. Non-Markovian SLE}

Very important results can be obtained in the model of interaction
$\hat {\cal L}_z$ highly localized in $\{z\}$-space, i.e. highly
localized jump rates $\hat k_z^d$ and $\hat k_z^n$. The most
interesting for our further discussion is the simplest variant of
this model, which in the case of discrete $\{z\}$-space is written
as
\begin{equation}\label{mark5}
\hat k_{\bf z}^d = \hat k_0 \hat P^{tt}_{\bf z}, \;\; \hat k_{\bf
z}^n = \hat k_0 \hat P^{nt}_{\bf z} \;\,\mbox{with} \;\, \hat
P^{\!jt}_{\bf z} = |{\bf z}_j\rangle\langle {\bf z}_t |
\end{equation}
and $j=t,n$. Here $\hat k_0$ is the matrix diagonal in
$\{x\}$-space, while $|{\bf z}_n \rangle$ and $|{\bf z}_t \rangle$
are the states $\{z\}$-space. The model (\ref{mark5}) approximates
${\cal L}(t)$-fluctuations by the set of jumps (blips) with the
relaxation effect of each jump described by $\hat k_0$-matrix. As to
properties of jumps statistics, they are determined by the
controlling operator $\hat \Lambda$ and the localized matrix $ \hat
P^{nt}_{\bf z}$ which implies that jumps occur as long as the system
appears at the state $|{\bf z}_t\rangle$ and each jump results in
the transition $|{\bf z}_t\rangle \to |{\bf z}_n\rangle$ in
$\{z\}$-space.

Evidently, for the case of continuum $\{z\}$-space in eq.
(\ref{mark5}) [with $\hat \Lambda$, for example, of type of that
given by eq. (\ref{mod1})] the term $\hat P^{nt}_{\bf z}$ should be
replaced by the corresponding $\delta$-function type one.

Within the model (\ref{mark5}) general formula (\ref{mark2_1a}) can
be represented in a more suitable CTRW-like form:
\begin{eqnarray}
\hat {\widetilde{\mathbb G}}(\epsilon)  &=&\langle {\bf z}_e| \hat
{G}(\epsilon)[1 - \hat P_x \hat k_0 \hat {\widetilde{G}}(
\epsilon)]^{-1}|{\bf z}_i\rangle\nonumber\\
&=& \hat {\widetilde{\mathbb P}}_{i}(\epsilon) + \hat
{\widetilde{\mathbb P}}_{n}(\epsilon)[1-\hat {\widetilde{\mathbb
W}}_{n} (\epsilon)]^{-1} \hat {\widetilde{\mathbb
W}}_{i}(\epsilon),\qquad \label{mark6}
\end{eqnarray}
where
\begin{equation}\label{mark7}
\hat {\widetilde{G}}(\epsilon) = (\epsilon + \hat L + \hat k_{\bf
z}^d + \hat \Lambda)^{-1},
\end{equation}
is the Green's function, that essentially determines two matrices:
the effective PDF-matrix $\hat {\widetilde{\mathbb W}}_{\!
j}(\epsilon)$ of sudden fluctuation jumps (analogs of renewals) and
the matrix $\hat {\widetilde{\mathbb P}}_{j}(\epsilon)$ of
probabilities not to observe any fluctuation until time $t$, playing
important role in CTRW-based theories \cite{Wei,Met,Hau}] [see eq.
(\ref{nmark1})]. These two matrices are expressed in terms of the
Green's function of the controlling process in $\{z\}$-space:
\begin{equation}\label{mark8}
\hat {\widetilde{\mathbb W}}_{\! j} (\epsilon) = \hat P_x \hat
{\widetilde{w}}_{\! j}(\epsilon), \; (j = i, n),
\end{equation}
in which
\begin{equation}\label{mark8a}
\hat {\widetilde{w}}_{\! j} (\epsilon) = \hat k_0\langle {\bf z}_t
|\hat G (\epsilon)  |{\bf z}_j \rangle = [1+\hat
\Phi_{\!j}(\epsilon)]^{-1},
\end{equation}
where
\begin{equation}\label{mark8ab}
\hat \Phi_{\!j}(\epsilon) = [\hat g_{tt}(\epsilon)\hat
g_{tj}^{-1}(\epsilon)-1] + [\hat k_0 \hat g_{tj}(\epsilon)]^{-1}
\end{equation}
with
\begin{equation}\label{mark8b}
\hat g_{t\!j}=\langle {\bf z}_t |\hat g (\epsilon)| {\bf z}_j
\rangle \;\;\mbox{and}\;\;\hat g(\epsilon) = (\epsilon + \hat L +
\hat \Lambda)^{-1},
\end{equation}
and
\begin{equation}\label{mark9}
\hat {\widetilde{\mathbb P}}_{j} (\epsilon) = \langle {\bf z}_e
|\hat G (\epsilon)  | {\bf z}_j \rangle = (\epsilon + \hat {L})^{-1}
[1 - \hat {\widetilde{w}}_{j} (\epsilon)].
\end{equation}
In eq. (\ref{mark9}) $\langle {\bf z}_e|$ is equilibrium ket-vector
in $\{ z\}$-space (with $\langle {\bf z}_e| \Lambda = 0$) defined by
formula similar to eq. (\ref{genf4a}) but with ${\bf x}$-vectors
replaced by those in $\{ z\}$-space.

Expression (\ref{mark6}), which in what follows will be called the
generalized non-Markovian SLE (in resolvent form), looks similar to
the simpler one derived earlier within the Markovian representation
\cite{Shu1}]. Moreover, for $\hat L = 0$ it reduces to that obtained
in refs. \cite{Shu1,Seki} in which the PDF matrices
\begin{eqnarray}
\hat {\widetilde{\mathbb W}}_{\! j} (\epsilon) &=& \hat
{\widetilde{\mathbb W}}_{\!0j} (\epsilon) = \hat P_x \hat
{\widetilde{W}}_{\!\!j}(\epsilon), \label{mark9c1}\\
\hat {\widetilde{\mathbb P}}_{j} (\epsilon) &=& \hat
{\widetilde{\mathbb P}}_{0j} (\epsilon) =  \epsilon^{\!-1} [1 - \hat
{\widetilde{W}}_{\!\!j} (\epsilon)],\;\;(j = i,n), \quad
\label{mark9c2},
\end{eqnarray}
where
\begin{equation} \label{mark9a}
\hat {\widetilde{W}}_{\!\!j} (\epsilon) = \!\int_0^{\infty} \!\!\!
dt \, \langle {\bf z}_e |\hat G_0 (t)  | {\bf z}_j \rangle e^{ -
\epsilon t} \equiv \langle {\bf z}_e |\hat {\widetilde{G}}_0
(\epsilon)  | {\bf z}_j \rangle
\end{equation}
is the PDF matrix in which $\hat G_{\!0} (t) = e^{-(\hat k_{\bf z}^d
+ \hat \Lambda)t}$ and $\hat {\widetilde{G}}_{\!0} (\epsilon) =
(\epsilon + \hat k_{\bf z}^d +  \hat \Lambda)^{-1}$, are the Green's
functions describing the stochastic process in $\{z\}$-space.
Equations (\ref{mark9c1}) and (\ref{mark9c2}) can be considered as a
generalized variant of relation (\ref{nmark4}).

Despite the similarity of obtained formulas with those known in the
CTRW theory \cite{Met,Hau}, there are, however, some important
specific features of the general expression (\ref{mark6}) as opposed
to the conventional CTRW formulas (\ref{nmark1})-(\ref{nmark4}) and
(\ref{mark9c1}), (\ref{mark9c2}).

1) The matrix
\begin{equation}\label{exten2}
\hat {\widetilde{\mathbb P}}_{\!j} (\epsilon) =  \hat \Omega^{-1}
(\epsilon) [1 - \hat k_0 \langle {\bf z}_t |(\hat \Omega (\epsilon)
+ \hat k_0 + \hat \Lambda)^{-1} | {\bf z}_j \rangle]
\end{equation}
in which $\hat \Omega (\epsilon) = \epsilon + \hat {L}$, can be
non-diagonal, unlike similar matrix in the conventional CTRW
approach (see below). Moreover, in general, elements $\hat {\mathbb
P}_{\!j_{ik}} (t)$ of the matrix $\hat {\mathbb P}_{\!j} (t)$ do not
satisfy the evident relation of type of $\hat {\mathbb P}_{\!j_{ik}}
(0) = 1$ which would allow one to interpret these elements as
probabilities. In addition, $\hat {\mathbb P}_{\!j} (t)$ can be
complex values, for example, in quantum processed for which the
elements of $\hat L$ are complex.

2) The relation between $\hat {\widetilde{\mathbb W}}_{\! j}
(\epsilon)$ and $\hat {\widetilde{\mathbb P}}_{j}(\epsilon)$ implied
by eqs. (\ref{mark8}) and (\ref{mark9}) does not coincide with that
between the PDF-matrix of waiting times and the matrix of
probabilities known in the CTRW theory. Strictly speaking, the
conventional relation [see eqs. (\ref{mark9c1}) and (\ref{mark9c2})]
is reproduced only in the evident case $\hat L = 0$.

3) Equation (\ref{mark6}) is of the form of CTRW expression however
with matrices $\hat {\widetilde{\mathbb W}}_{\! j} (\epsilon)$ and
$\hat {\widetilde{\mathbb P}}_{\!j} (\epsilon)$ strongly modified by
the dynamic operator $\hat L$ [in contrast to the non-Markovian SLE
appealing to the conventional PDF matrices $\hat {\mathbb W}_{0\!j}
(t)$ and $\hat {\mathbb P}_{0\!j} (t)$ \cite{Shu1} (see eq.
(\ref{mark12}))].

The above-mentioned effects of dynamic evolution can result in
significant change of the time dependencies $\hat {\mathbb W}_{0\!j}
(t)$ and $\hat {\mathbb P}_{0\!j} (t)$ thus leading to the strong
change of the kinetics of relaxation processes under study.

\subsubsection{CTRW-based non-Markovian SLE.}

The simplified variant of formula (\ref{mark6}) can be obtained in
the special case of matrix $\hat {\widetilde{W}}_{\! j} (\epsilon)$
diagonal in $\{x\}$-space, which is realized when the matrices $\hat
\Lambda$, $\hat L$, and $\hat k_0$ commute with each other: $[\hat
\Lambda,\hat L] = [\hat L,\hat k_0] = 0$. In this case one can
obtain the representation for $\hat {\widetilde{\mathbb W}}_{\! j}
(\epsilon)$ in terms of the Laplace transform of the conventional
PDF-matrix of fluctuation blips $\hat {\widetilde{\mathbb
W}}_{\!0\!j}$ [see eqs. (\ref{mark9c1})) and (\ref{mark9a}))]:
\begin{equation}\label{mark12}
\hat {\widetilde{\mathbb W}}_{\! j} (\epsilon) = \hat
{\widetilde{\mathbb W}}_{\! 0\!j}(\hat \Omega(\epsilon)) = \hat P_x
\!\int_0^{\infty} \!\!\! dt \, {\hat W}_{\!j}(t) e^{ - \hat
\Omega(\epsilon) t},
\end{equation}
where $\hat \Omega (\epsilon) = \epsilon + \hat L$.

The simplified non-Markovian SLE (\ref{mark12}), is nevertheless
more general than the variant of this equation in which $\hat
{\widetilde{W}}_{\!0\! j} (\epsilon)$ independent of $x$-coordinate
(i.e. $\hat {\widetilde{W}}_{\!0\!j} (\epsilon)$ proportional to the
unity matrix in $\{x\}$-space) \cite{Bursh}.

Noteworthy is that, in general, if $\hat \Lambda$, $\hat L$, and
$\hat k_z$ do not commute with each other, the representation
(\ref{mark12}) is not valid and one should use the original
expression (\ref{mark6}).

\section{Extensions of CTRW models}

The proposed Markovian representation, based on description of CTRWs
as MP affected MPs, enables one to significantly extend the CTRW
approach and the non-Markovian SLE. It reduces the treatment of
system evolution to averaging exponential functional over Markovian
fluctuations and offers the expression of CTRW equations in terms of
multidimensional Markovian SLE.

There are some straightforward extensions which do not need detailed
analysis. For example, the evident variant of extension is the model
of several highly localized jump states in $\{z\}$-space $\langle
{\bf z}_{t_i} |$ and $|{\bf z}_{n_i} \rangle$ ($1 \leq i \leq i_z,
\, i_z > 1$) in which $k_z = \mbox{$\sum_j k_{0}^{(\!j)}\hat
P_{nt}^{(\!j)}$}$ with $\hat P_{nt}^{(\!j)} = |{\bf
z}_n\rangle\langle {\bf z}_t |$. In this model the general formula
(\ref{mark2_1a}) is also simplified by reducing the problem to
solving the linear equation for the matrix $\hat {\bf G}$ with
elements $\hat {\bf G}_{{ij}}(\epsilon) = \langle {\bf z}_{t_i} |
\hat G (\epsilon) | {\bf z}_{n_j}\rangle$. Unfortunately, in this
model the expression for $\hat {\widetilde{\mathbb G}}$ is fairly
cumbersome and not quite suitable for applications.

In this section we will mainly discuss less evident and more general
extensions which allow for describing effects of Markovian and
non-Markovian fluctuating interactions on kinetics of relaxation in
CTRW-like systems, i.e. (in our brief terminology) MP and CTRW
affected CTRWs. In the Markovian representation the problem is still
reduced to the analysis of MP affected MPs though in
multidimensional space.

Because of large number of parameters in this type of
multidimensional MPs it is practically senseless to discuss the
problem in general. For this reason, here we will restrict ourselves
to consideration of two modifications and extensions of the CTRW
approach interesting for theoretical analysis and for applications.

\subsection{CTRW-control. Coupled Markovian processes.}

\subsubsection{General results}

So far in our analysis we have discussed the Markovian model for
$\hat {\cal L} (t)$ fluctuations which suggests that the controlling
stochastic process $z(t)$, responsible for fluctuations of the jump
rates $\hat k_{{\bf z}(t)}^{d,n}$ [see eq. (\ref{genf4})], is
Markovian.

Here we will consider the extension of the Markovian model based on
the assumption that the controlling process $z(t)$ is represented as
a sequence (cascade) of controlling Markovian processes in the
multidimensional space $\{{\bf Z}_1 \} = \{z_1, z_2, \dots, z_N\}$
with evolution in each space $\{z_q\}$ being controlled by the
process in $\{z_{q+1}\}$-space whose mechanism is similar to that
responsible for CTRW-type motion in $\{x\}$-space as described above
in Sec. III.B. This mechanism implies that stochastic jumps in
$\{z_q\}$-space happen as long as the system appears in the
transition state $|{\bf z}_{q+1}^t\rangle$ (in $\{z_{q+1}\}$-space).
As compared to the mechanism presented in Sec III.B, however, here,
for simplicity, we will assume that jumps are not accompanied by the
change of state in $\{z_{q+1}\}$-space, i.e. the final state $|{\bf
z}_{q+1}^n\rangle = |{\bf z}_{q+1}^t\rangle$. For the same reason we
will also assume that initial state $|{\bf z}_{q}^{i}\rangle$ of
evolution in $\{z_{q}\}$-subspace coincides with the transition
state: $|{\bf z}_{q}^{i}\rangle = |{\bf z}_{q}^t\rangle$ ($1 \leq q
\leq N$).

The above-formulated model of cascaded controlling processes is
described by the following jump operators in spaces $\{{\bf Z}_q \}
= \{z_q, z_2, \dots, z_N\} \: (q \geq 1)$
\begin{equation}
\hat {\cal L}_q = \sum\nolimits_{i=q}^N \!\hat \Lambda_i \hat
P^{tt}_{\!i+1}, \;\;\mbox{where}\;\; \hat \Lambda_i  = (1\!-\! \hat
{\cal P}_{z_i}) \hat k_{0_i} \label{ctrw0}
\end{equation}
with
\begin{equation}
\hat P^{tt}_{i} = |{\bf z}_{i}^t\rangle \langle {\bf z}_{i}^t|.
\label{ctrw01}
\end{equation}
Here $\hat {\cal P}_{z_i}$ is the matrix of distribution functions
of jump lengths (in $\{z_i\}$-space) and $\hat k_{0_i}$ is the
matrix of jump rates diagonal in $\{z_i\}$-space.

Similar to the model discussed in Sec. III.B, the control of motion
in $\{x\}$-space will be described by $z$-dependence of the operator
${\cal L}$:
\begin{equation}
\hat {\cal L}_{\bf z} = (1-\hat {\cal P}_{x}) \hat k_{0} \hat
P^{tt}_{1}. \label{ctrw1}
\end{equation}

Noteworthy is that in accordance with obtained results the model
(\ref{ctrw0})-(\ref{ctrw1}) can be considered as a generalized
variant of CTRW-type models for the controlling process.

Recall that the problem under study reduces to evaluating the
evolution operator ${\hat {\mathbb{G}}}(t) $ defined in eq.
(\ref{genf4_0}). In the proposed model of cascaded controlling
processes this operator can be found in analytical form.

Formula for ${\hat {\mathbb{G}}}(t) $ can be obtained with the use
of the general expression (\ref{mark6})
\begin{eqnarray}\label{ctrw2}
\hat {\widetilde{\mathbb G}}(\epsilon)  &=& \epsilon^{-1}[1 - \hat w
(\epsilon)][1-\hat P_x \hat w (\epsilon)]^{-1} \label{ctrw2a}\\
&=&\epsilon^{-1} \hat \Phi (\epsilon)[\hat \Phi (\epsilon)+\hat
{\cal L}_x\hat k_0^{-1}]^{-1} \label{ctrw2b}
\end{eqnarray}
in which $\hat {\cal L}_x = (1 - \hat P_x)\hat k_0$,
\begin{equation}\label{ctrw3}
\hat w (\epsilon) = [1+ \hat \Phi (\epsilon)]^{-1}
\;\,\mbox{with}\;\,\hat \Phi (\epsilon) = \hat k_{0}^{-1}\hat
\phi_{1}(\epsilon \!+\! \hat {\cal L}_2)
\end{equation}
and
\begin{equation}\label{ctrw4}
\hat \phi_1 (\epsilon)= \langle {\bf z}_1^t|\hat g_1(\epsilon)|{\bf
z}_1^t\rangle^{-1} = \langle {\bf z}_1^t|(\epsilon + \hat
\Lambda_1)^{-1}|{\bf z}_1^t\rangle^{-1} .
\end{equation}

The expressions (\ref{ctrw2})-(\ref{ctrw4}) relates the evolution
operator $\hat {\widetilde{\mathbb G}}(\epsilon) $ of the total
system in the combined space $\{x \otimes {\bf z}\}$ to that $\hat
g_1(\epsilon) = (\epsilon + \hat \Lambda_1)^{-1}$ in the subspace
$\{{\bf Z}_1\}$.

The procedure presented above allows one to express $\hat \phi_1
(\epsilon\!+\!{\cal L}_2)$ in terms of the evolution operator $\hat
g_{2}(\epsilon)$ in the reduced subspace $\{{\bf Z}_2\}$. The
expressions similar to eqs. (\ref{ctrw2})- (\ref{ctrw4}), as applied
to the operator $\hat g^{tt}_{1}(\epsilon)$, yield
\begin{equation}\label{ctrw5}
\hat \phi_{1}(\epsilon) = \langle {\bf z}_2^t|[\hat \phi_2 (\epsilon
+ \hat {\cal L}_3) + \hat \Lambda_2]^{-1}|{\bf z}_2^t\rangle^{-1}
\end{equation}
with
\begin{equation}\label{ctrw6}
\hat \phi_2 (\epsilon) = \hat g^{tt}_{2}(\epsilon)^{-1}
\end{equation}
and
\begin{equation}\label{ctrw6-1}
\hat g^{tt}_{2}(\epsilon) = \langle {\bf z}_2^t|\hat
g_2(\epsilon)|{\bf z}_2^t\rangle = \langle {\bf z}_2^t|(\epsilon +
\hat \Lambda_2)^{-1}|{\bf z}_2^t\rangle .
\end{equation}

By continuing the proposed procedure one gets the expression for the
function $\hat \Phi(\epsilon)$ in terms of $\hat
g_{q}^{tt}(\epsilon)$ (with $q \geq 1$), and therefore in terms of
functions
\begin{equation}\label{ctrw7}
\hat \phi_q (\epsilon) =  \langle {\bf z}_q^t|(\epsilon +
\hat\Lambda_q)^{-1}|{\bf z}_q^t\rangle^{-1},
\end{equation}
which describe memory effects in the system as a result of Markovian
motion in $\{{ z}_q\}$-subspace, i.e. without controlling
interaction with other subspaces $\{{z}_i\}$ with $i > q$:
\begin{equation}\label{ctrw8}
\hat \Phi (\epsilon) = \hat k_0^{-1}\hat \phi_1 (\hat \phi_2(\hat
\phi_3(\dots))) (\epsilon).
\end{equation}

\subsubsection{Examples}

To illustrate the obtained results we will consider two limiting
examples of two-state exponential and anomalously slow inverse power
type controlling processes.

\paragraph{Two-state controlling processes.}

The exponential two-model for controlling processes corresponds to
the simple expression for the evolution operator
\begin{eqnarray}\label{ctrw9}
\hat \phi_q (\epsilon) &\sim& \langle {\bf z}_q^t|(\epsilon +
\hat\Lambda_q)^{-1}|{\bf z}_q^t\rangle^{-1} \nonumber \\
& = & \epsilon_{q}/\epsilon - \kappa_q /(\epsilon  + \xi_q),
\end{eqnarray}
where $\epsilon_{q}, \: \kappa_q,$ and $\xi_q$ are some constant
parameters. Obviously, formula (\ref{ctrw8}) with  $\hat \phi_q
(\epsilon)$ presented in eq. (\ref{ctrw9}) gives the expression for
$\hat \Phi (\epsilon)$ of type of continued fraction which predicts
multiexponential behavior of the PDF matrix $\hat w (t)$.

\paragraph{Anomalously slow controlling processes.}

Another very important model of controlling processes describes
anomalously slow inverse-power type behavior of the PDF $\hat w_1
(t)$. This model is realized by taking a weak fractional-power type
dependence of $\hat \phi_q (\epsilon)$ \cite{Met}:

\begin{equation}\label{ctrw10}
\hat \phi_q (\epsilon) = w_q(\epsilon/w_q)^{\alpha_q}, \;\;\alpha_q
< 1,
\end{equation}
where $w_q = \zeta_q k_{0_q}$ and $\zeta_q \sim 1$. Substitution of
eq. (\ref{ctrw10}) into formula (\ref{ctrw8}) yields
\begin{equation}\label{ctrw11}
\hat \Phi (\epsilon) = R_0(\epsilon/w_N)^{\alpha_0}, \;\;\mbox{with
}\alpha_0 = \prod\nolimits_{q=1}^N \!\alpha_q < 1
\end{equation}
and $R_0 =\prod_{q=1}^N (w_q/w_{q-1})^{\gamma_q},$ where $\gamma_q =
\prod_{i=1}^q \alpha_{i}$.

Note that in the simplest case of identical rates $w_q$: $w_q =
\bar{w} ,\: (q \geq 1),$ when $R_0 = 1$ and $w_N \approx \bar{w}$,
formula (\ref{ctrw11}) is represented in a simple form $\hat \Phi
(\epsilon) \approx (\epsilon/\bar{w})^{\alpha_0}$.

\subsection{Fluctuating CTRW-jumps}

\subsubsection{General formulas}

Another interesting problem in the CTRW theory, which can thoroughly
be analyzed with the use of the proposed Markovian representation,
concerns the kinetics of CTRW processes governed by stochastically
fluctuating PDF matrices $\hat {{\mathbb W}}_{\beta}(t)$ of jumps
(renewals) or corresponding matrices $\hat {w}_{\beta}(t)$ (here
$\beta$ is the number of jump).

The fact is that in conventional CTRW theories the PDF $\hat
{W}_{\beta} (t)$ and the probability $\hat {P}_{\beta}(t)$ are
assumed to be the same for all renewals $\beta > 1$ except the first
one ($\beta = 0$) and fixed functions of time [eq. (\ref{nmark1})].
In this section we will discuss the extension of CTRW processes in
which the fluctuation jump kinetics is controlled PDF and
probability matrices whose functional form fluctuates leading, in
particular, to the difference of $\hat {{\mathbb W}}_{\beta} (t)$
for different $\beta$.

The main difficulty in modeling fluctuating PDF matrices consists in
necessity to take into account the normalization condition
$\int_0^{\infty} \!dt\, W_{\beta}(t) = \hat {\widetilde
{W}}_{\beta}(\epsilon = 0) = 1$, which ensures the population
conservation in the process.

The Markovian representation, discussed in Sec. III, allows us to
make the above problem tractable. Within this representation the
fluctuations of PDF functions $\hat {{w}}_{\beta}(t)$ are assumed to
be determined by the Markovian controlling process which is governed
by the fluctuating operator $\hat \Lambda$. The representation
offers quite natural and fairly simple way of description of
fluctuating $\hat \Lambda (t)$ by suggesting this operator to depend
on the parameter ${\bf y}(t)$ which undergoes stochastic Markovian
fluctuations. In this model the validity of the normalization
relation for $\hat {w}_{\beta}(t)$ can be ensured by taking proper
form of the operator $\hat \Lambda (t)$. It is clear from eq.
(\ref{mark1_1}) that this relation will be fulfilled if during
evolution in $\{z\}$-space, described by fluctuating $\hat \Lambda
(t)$, the population is conserved, i.e. eq. (\ref{mark1_1}) can be
represented in the form $\dot \varphi = - (\nabla_z \cdot {\bf J}_z
(\varphi,t))$, where ${\bf J}_z (\varphi,t)$ is the fluctuating flux
in $\{z\}$-space. For instance, the population is certainly
conserved in processes governed the Smoluchowski operator
(\ref{mod1}). More clearly this fact will be illustrated below with
some examples.

In general in the Markovian model proposed above, the PDF $\sigma
(y|t)$ representing $y(t)$-fluctuation process satisfies equation of
type of (\ref{mark1_1}):
\begin{equation}\label{ctrw13}
\dot \sigma = - \hat {\Lambda}_y \sigma \;\;\; \mbox{with} \;\;\;
\sigma ({\bf y},0) = \sigma_i({\bf y}),
\end{equation}
where $\hat \Lambda_y$ is the operator responsible for the evolution
in $\{y\}$-space.

Similar to the case of $\hat {\cal L}(t)$ fluctuations caused by
Markovian fluctuating ${\bf z}(t)$-parameter, the consideration of
the effect of ${\bf y}(t)$-fluctuations reduces to analyzing the SLE
for the evolution operator $\hat {{\mathbb G}}_{\bf r}(t)$ in the
combined space $\{{\bf r}\}=\{x \otimes z \otimes y\}$:
\begin{equation}
\dot {\hat {\mathbb{G}}}_{\bf r} = -(\hat {\cal L} + \hat
{\Lambda}_z + \hat {\Lambda}_y) \hat {\mathbb{G }}_{\bf r}
\;\:\mbox{with} \;\;\hat {\mathbb{G}}_{\bf r} ({\bf r},{\bf r}_i|0)
= \delta_{{\bf r} {\bf r}_i}. \label{ctrw14}
\end{equation}
so that
\begin{equation}
{\hat {\widetilde{\mathbb{G}}}} = \langle {\bf y}_e| \langle {\bf
z}_e|(\epsilon + \hat {\cal L} + \hat {\Lambda}_y + \hat
{\Lambda}_z)^{-1}|{\bf z}_i\rangle |{\bf y}_i \rangle,
 \label{ctrw15}
\end{equation}
where $|{\bf y}_i\rangle \equiv \sigma_i({\bf y})$ is the initial
${\bf y}(t)$-state in bra/ket notation and
\begin{equation}
\langle \hat {\widetilde {U}} \rangle = \langle (\epsilon + \hat
{\cal L} + \hat {\Lambda}_y + \hat {\Lambda}_z)^{-1}\rangle_{\bf r},
 \label{ctrw16}
\end{equation}
In eq. (\ref{ctrw14}) both the jump operator $\hat {\cal L}$ and the
controlling operator $\Lambda_z$ can, in principle, depend on
variable $y$ as a parameter: $\hat {\cal L} \equiv \hat {\cal
L}_{z}^y$ and $\Lambda_z \equiv \Lambda_z^y$.

Just these dependencies of $\hat {\cal L}$ and $\Lambda_z$ on $y$
model the effect of ${\bf y}(t)$ evolution [i.e. fluctuations of PDF
functions $\hat {{w}}_j(t)$] on the kinetics of processes under
study. In general, both dependencies can strongly contribute to the
effect, however in this work we will concentrate on the analysis of
the effect of $\hat {\cal L}_{z}^y$-dependence only, assuming that
$\Lambda_z$ is independent of ${\bf y}$.

In the Markovian model of $\hat {{w}}_j(t)$-fluctuations the problem
reduces to the analysis of the SLE (\ref{ctrw14}) which is similar
to eq. (\ref{mark2_1}) considered above in detail. In general, eq.
(\ref{ctrw14}) can be solved only numerically. In this work we will
restrict ourselves to discussing some most important specific
features of manifestation of $\hat {{w}}_j(t)$-fluctuations in the
kinetics of the process within the simple analytically solvable
variant of the proposed model.

\subsubsection{Localized rate of fluctuation jumps}

The variant of the Markovian model which allows for analytical
analysis of the problem can be considered as an extension of the
approximation of highly localized jump rates discussed above. Here
we assume high localization of fluctuation jump rates in both spaces
$\{z\}$ and $\{y\}$, i.e. in the combined space $\{z \otimes y\}$.
In this limit the expression for $\hat{\cal L}_z^y $-dependence is
given by
\begin{equation}\label{ctrw17}
\hat{\cal L}_z^y = (1\!-\!\hat P_x)\hat k_0 \hat P^{tt}_{\!yz}
\;\;\mbox{with}\;\; P^{tt}_{yz} = |{\bf y}_t{\bf z}_t\rangle \langle
{\bf y}_t{\bf z}_t|,
\end{equation}
where $|{\bf y}_t{\bf z}_t\rangle = |{\bf y}_t\rangle |{\bf
z}_t\rangle $ is the jump state in $\{z \otimes y\}$-space which,
for simplicity, is suggested to remain unchanged after jumps.

In the model (\ref{ctrw17}) $y(t)$-fluctuations are represented as a
set of jumps (blips). The statistics of jumps is characterized by
the PDF $w_y (t)$ of waiting times of jumps whose Laplace transform
can be written as:
\begin{equation}\label{ctrw17a}
\hat{\widetilde{w}}_{y} (\epsilon) = [1+ \hat
\Phi_{y}(\epsilon)]^{-1},
\end{equation}
where
\begin{equation}\label{ctrw17b}
\hat \Phi_{y}(\epsilon) = [\hat k_{0}\langle {\bf y}_t|(\epsilon +
\hat \Lambda_y)^{-1}|{\bf y}_t\rangle]^{-1}.
\end{equation}

It is easily seen that from mathematical point of view the proposed
model is close that formulated in Sec. III. The difference consists
only in a larger dimensionality of the space of the controlling
process ($\{z \otimes y\}$-space instead of $\{z\}$-one in the model
considered in Sec. III) and in the corresponding difference in the
form of the operator $\Lambda$ which describes the controlling
process ($\Lambda_z + \Lambda_y$ instead of $\Lambda_z$).

This similarity enables us to use the results obtained above. It
follows from these results that the model (\ref{ctrw17}) predicts
the CTRW expression for the evolution operator ${\hat
{\widetilde{\mathbb{G}}}}$ of type of eq. (\ref{mark6}):
\begin{equation}\label{ctrw18}
\hat {\widetilde{\mathbb G}}(\epsilon)  = \hat {\widetilde{\mathbb
P}}_{i}(\epsilon) + \hat {\widetilde{\mathbb
P}}_{t}(\epsilon)[1-\hat {\widetilde{\mathbb W}}_{t}
(\epsilon)]^{-1} \hat {\widetilde{\mathbb W}}_{i}(\epsilon),
\end{equation}
where
\begin{equation}\label{ctrw19}
\hat {\widetilde{\mathbb W}}_{\! j} (\epsilon) = \hat P_x \hat
{\widetilde{w}}_{f_j}(\epsilon)  \;\;\mbox{and} \;\;\hat
{\widetilde{\mathbb P}}_{\!j} (\epsilon) = \epsilon^{-1} [1 - \hat
{\widetilde{w}}_{f_j} (\epsilon)],
\end{equation}
$(j = i, t)$, and
\begin{equation}\label{ctrw20}
\hat{\widetilde{w}}_{\!f_j} (\epsilon) = [1+ \hat
\Phi_{f_j}(\epsilon)]^{-1},
\end{equation}
however, with matrices
\begin{equation}\label{ctrw21}
\hat \Phi_{f_j}(\epsilon) = [\hat k_{0}\langle {\bf y}_t{\bf
z}_t|(\epsilon + \hat \Lambda_y + \hat \Lambda_z)^{-1}|{\bf y}_t{\bf
z}_j\rangle]^{-1}.
\end{equation}
whose behavior essentially depends on specific features of
stochastic evolution of ${\bf y}(t)$.

This expression makes it possible to easily analyze some specific
features of the effect of ${\bf y}(t)$-fluctuations. First of all,
it is of special interest to discuss two natural limits: slow and
fast ${\bf y}(t)$-fluctuations.

{\it a. The limit of slow $y(t)$- and $\hat {{w}}_j
(t)$-fluctuation.} \,In the limit of slow fluctuations $y(t)$, when
the characteristic time of evolution in $\{y\}$-space, $\tau_y$, is
larger than that of the PDFs $\hat {{w}}_j(t)$, $\tau_w$ : $\tau_y
\gg \tau_w$, one can neglect the term $\hat \Lambda_y$ in eqs.
(\ref{ctrw14})-(\ref{ctrw16}) thus arriving at the expression for
${\hat {\mathbb{G}}}$ and $\langle \hat {\widetilde {U}} \rangle $
of type of those obtained in CTRW approach (see Sec. III) but
averaged over $y$-distribution $|\sigma_i({\bf y})\rangle \equiv
|{\bf y}_i\rangle$. In the short range model considered in this
limit one gets the expression (\ref{ctrw18})-(\ref{ctrw21}) for
$\hat {\widetilde{\mathbb G}}(\epsilon)$ with:
\begin{equation}\label{ctrw22}
\hat \Phi_{f_j}(\epsilon) \approx [\hat k_{0}\langle {\bf y}_t|{\bf
y}_j\rangle\langle {\bf z}_t|(\epsilon + \hat \Lambda_z)^{-1}|{\bf
z}_j\rangle]^{-1}.
\end{equation}

{\it b. The limit of fast $y(t)$- and  $\hat {{w}}_j
(t)$-fluctuations.}\, In the opposite limit of fast fluctuations
${\bf y}(t)$, corresponding to $\tau_y \ll \tau_w$, the SLE
(\ref{ctrw14}) still yields the CTRW-like formula for the evolution
operator ${\hat {\mathbb{G}}}$. In this formula, however, the
matrices $\hat {\widetilde{\mathbb W}}_{\! j} (\epsilon)$ and $\hat
{\widetilde{\mathbb P}}_{\! j} (\epsilon)$ are expressed in terms of
the PDF matrix $\hat {{w}}_j (t)$ obtained by means of relations
(\ref{mark8})-(\ref{mark9}) which are evaluated with transition
matrices $\hat k_z^d ({\bf y})$ ana $\hat k_z^n ({\bf y})$ [see eq.
(\ref{mark5})] averaged over the equilibrium $y$-distribution
$\sigma_e ({\bf y}) \equiv |{\bf y}_e\rangle$. In the fast
fluctuation limit the short range model results in the expression
(\ref{ctrw18})-(\ref{ctrw21}) for $\hat {\widetilde{\mathbb
G}}(\epsilon)$ with:
\begin{equation}\label{ctrw23}
\hat \Phi_{f_j}(\epsilon) \approx [\hat k_{0}\langle {\bf y}_t|{\bf
y}_e\rangle\langle {\bf z}_t|(\epsilon + \hat \Lambda_z)^{-1}|{\bf
z}_j\rangle]^{-1},
\end{equation}
where $|{\bf y}_e\rangle$ is the equilibrium state in $\{y\}$-space.

\medskip

It is seen from formulas (\ref{ctrw22}) and (\ref{ctrw23}) that in
the considered simple model of localized jump rates the kinetics of
the process is similar in both limits. The difference is only in the
characteristic parameters. However, in general, in the case of
delocalized jump rates, i.e. delocalized dependence of ${\cal
L}_z^y$ on ${\bf y}$, the kinetics in these two limits is, of
course, different though the asymptotic behavior at long times is
expected to be similar. Below we will discuss the effect of
delocalization of this dependence (see next Section).

In addition, strictly speaking, in the above analysis of two limits
we have assumed that ${\bf y}(t)$-fluctuations are stationary which
implies the existence of the equilibrium state $|{\bf y}_e\rangle$
of the operator $\Lambda_y$. It is of special interest to study the
manifestation of non-stationary ${\bf z}(t)$-fluctuations which can
be fairly strong especially in the case anomalously slow
fluctuations \cite{Met}.

\subsubsection{Anomalous $y(t)$-fluctuations}

Here we will analyze in detail the effect of anomalously slowly
fluctuating coordinate ${\bf y} (t)$. In the Markovian
representation this type of fluctuations can be described with the
use of the free diffusion model for $y(t)$-evolution discussed in
Sec. III.B1. In this model the operator $\Lambda_y$ is given by
\begin{equation}\label{ctrw24}
\hat \Lambda_y = -  y^{1-n_y}\nabla_y(D_y y^{n_y-1}\nabla_y),
\end{equation}
where $\nabla_y = \partial/\partial y\,$ and $\,n_y\,$ is the
$\{y\}$-space dimensionality (it is assumed that $n_y \leq 2$).

The model (\ref{ctrw24}) predicts the anomalous long-time-tailed
dependence of the characteristic PDF $w_y (t)$ [see formulas
(\ref{ctrw17a}) and (\ref{ctrw17b})]:
\begin{equation}\label{ctrw25}
w_y (t) \sim 1/t^{1+n_y/2}.
\end{equation}
These anomalous fluctuations of PDFs $w_{\beta}(t)$ of CTRW-jumps
very strongly affect the kinetics of CTRW process. The effect can
clearly be revealed by analyzing the behavior of the matrices
$\Phi_{f_j} (\epsilon)$ [see eq. (\ref{ctrw21})]. Of special
interest and importance is the asymptotic behavior of $\Phi_{f_j}
(\epsilon)$ at $\epsilon \to 0$ which determines the long time
asymptotic behavior of the process.

To demonstrate the specific features of the small-$\epsilon$
behavior of $\Phi_{f_j} (\epsilon)$ we will consider two examples of
controlling processes in $\{z\}$-space: exponential (Poissonian) and
anomalous long-time-tailed. Both these examples can properly be
treated within the diffusion model.

{\it a. Poissonian ${\bf z}(t)$ controlling processes.} The
Poissonian-like controlling processes can be described by the model
of diffusive motion of the Brownian particle confined within the
well in $\{z\}$-space. In this model at small $\epsilon$,  which
correspond to times longer than the time $\tau_z$ of diffusive
relaxation within the well $U(z)$ in $\{z\}$-space, the behavior of
$\Phi_{f_i} (\epsilon)$ is determined by free diffusion in
$\{y\}$-space and therefore
\begin{equation}\label{ctrw26}
\Phi_{f_j} (\epsilon) \sim
\epsilon^{n_y/2}\;\;\mbox{and}\;\;w_{f_j}(t)\sim 1/t^{1+n_y/2},
\end{equation}
$(j = i, t)$. This result shows that originally normal CTRW
processes become anomalous as a result of anomalous
$w_j(t)$-fluctuations and the resultant behavior of $w_j(t) \equiv
w_{f_j}(t)$ coincides with that of $y(t)$-fluctuations, i.e.
anomalous $y(t)$-fluctuations strongly modify the kinetics of normal
CTRW processes.

{\it b. Long-time-tailed ${\bf z}(t)$ controlling processes.} The
long-time tailed case is represented by the free diffusion model in
$\{z\}$-space [$U(z) = 0$]. In this model the small
$\epsilon$-behavior of $\Phi_{f_i} (\epsilon)$ is determined by free
diffusion in the total $\{y\otimes z\}$-space of dimensionality $n_y
+ n_z$. This means that in the model of localized jump rates
\begin{equation}\label{ctrw27}
\Phi_{f_j} (\epsilon) \sim
\epsilon^{n_{yz}/2}\;\;\mbox{and}\;\;w_{f_j}(t)\sim
1/t^{1+n_{yz}/2},
\end{equation}
where $n_{yz} = n_y + n_z$. Formula (\ref{ctrw27}) demonstrates the
strong effect of $w_j (t)$-fluctuations on the kinetics of the
originally anomalous CTRW processes. Noteworthy is that $w_j
(t)$-fluctuations result in the increase of the anomaly parameter
$\alpha$ which determines the long time behavior of $w_j (t)$: $w_j
(t) \sim 1/t^{1+\alpha}$.

It is interesting to note that the effect of fluctuations can lead
to the crucial change of fluctuation jump statistics. The fact is
that usually CTRW theories assume that $\int_0^{\infty} dt \,
W_{\beta} (t) = 1$. This relation ensures conservation of
normalization (or population) in CTRW processes. In the free
diffusion approximation for the controlling process this
conservation relation is fulfilled if the dimensionality of the
space is small, when the statistics of reoccurrences in the jump
state is recurrent. That is why we have assumed $n_y < 2$ and $n_z <
2$. These two inequalities, however, do not warrantee the same
inequality for  $n_{yz} = n_y + n_z$. In principle, one can get
$n_{yz} > 2$ and in this case the the statistics of reoccurrences
becomes transient which means that $\int_0^{\infty} dt \, w_j (t <
1$. Such CTRW processes with decay are known in the probability
theory although are not applied widely \cite{Fel}.

\section{Discussion}

The results obtained in this work have demonstrated that the
Markovian representation is very useful for the analysis and
extensions of the CTRW approach. In this section we will discuss
some specific features of the proposed method and obtained results.

But first we would like to emphasize the important point concerning
the relation of this representation to conventional approaches
applied in the theory of CTRW processes. The fact is that the
Markovian representation can be considered as a convenient method of
realizing subordination (in a fairly general form) which is
conventionally used as a basis for formulation of the CTRW approach
\cite{Fel,Wei,Sok2}. Within this representation the subordination is
associated with the controlling process in $\{{\bf z}\}$-space (Sec.
III.B). The important advantage of the proposed realization consists
in simplification of the description of non-Markovian kinetic
problems by reducing the treatment to manipulations with linear
operators. Moreover, in many cases the representation allows for
deep understanding and modeling of real stochastic processes in
clear physical terms.

\subsection{Applicability of CTRW approach}

The Markovian equation (\ref{mark2_1}), applied to deriving the
Markovian representation, is of course more general than the CTRW
approach and therefore is quite suitable for the analysis of
applicability conditions of this approach. Here we will present some
comments on this point.

The problem reduces to analyzing the statistics of fluctuation jumps
controlled by the stochastic process in $\{z\}$-space which is
determined by the evolution operator $ \hat G (\epsilon) = (\epsilon
+ \hat k_z^d + \hat \Lambda)^{-1}$,  where $\hat k_z^d $ is the
jump-rate matrix diagonal in $\{x\}$- and $\{z\}$-spaces [see eq.
(\ref{mark7})].

In the limit of high localization of jump-rate matrix $\hat k_z^{d}
$ (in $\{z\}$-space) defined by eq. (\ref{mark5}) the Markovian
equation (\ref{mark2_1}) leads to the evolution operator $\hat
{\widetilde{\mathbb G}} (\epsilon)$ corresponding to the CTRW
approach. In the case of extended $\hat k_z^d $, however, the CTRW
approach is not valid, strictly speaking.

To find the conditions which can ensure the applicability of the
CTRW approach we will consider above-proposed simple and fairly
flexible diffusion model for the controlling operator $\hat \Lambda$
[eq. (\ref{mod1})].

1) {\it Poissonian-like statistics.} Within the diffusion model the
Poissonian-like jump statistics is described by the evolution
operator $\hat G (\epsilon)$ with the Smoluchowski operator $\hat
\Lambda$, in which the potential $u(z)$ is of type of infinitely
deep potential well, say, of width $a_u$, i.e. $u (a_u) \sim 1$. At
long times the proposed model predicts exponentially decreasing
function $w_n (t)$ with the character time $\tau_n \sim a_u^2/D_z$.

It is easily seen that in this model the limit of high localization
is realized for $a_k << a_u$, where $a_k$ is the characteristic
width of the function $\hat k_z^d $.

2) {\it Anomalous long-time tailed statistics.} The more interesting
case of long-time tailed jump statistics is represented by the free
diffusion variant of the diffusion model ($u(z) = 0$), which in the
limit of highly localized $\hat k_z^d $ predicts the PDF  $w_j (t)
\sim 1/t^{1 + n_z/2}$, where $n_z$ is the dimensionality of
$\{z\}$-space.

In the case of long-time tailed statistics the condition of
applicability of the CTRW approach (applicability of the
approximation of highly localized jump rates) is less trivial than
that formulated above for Poissonian-like statistics. The fact is
that in the absence of the potential one needs to compare the
characteristic width of $\hat k_{ z}^d $ in $\{z\}$-space with that
of the PDF $\varphi (z,t)$ of the controlling process [see eq.
(\ref{mark1_1})] which yield the condition $a_k^2 << D_z t$.

It is important to note that to satisfy this condition the function
$\hat k_{z}^d $ does not need to be very short range. It can be
shown that, for example, in the case $\hat k_{\bf z}^d \sim 1/z^m$
the proper parameter $a_k$ can be introduced for $m > 3$
\cite{Shu4}. In addition in the case $m > 3$ the long time behavior
of the PDF is shown to be identical to that for really short range
$\hat k_z^d$ with the corresponding size $a_k$. This means that for
$m > 3$ at long times $a_k << D_z t$ the CTRW approach is quite
applicable.

This brief analysis shows that {\it in the large class of anomalous
non-Markovian models the long time behavior of the evolution
operator $\hat {{\mathbb G}} (t)$ is correctly described by the CTRW
approach.}

\subsection{Extended CTRW approaches}

In Sec. IV we have studied most general features of two extensions:
CTRW with cascaded controlling processes and CTRW processes with
fluctuating PDFs. Below we will discuss in detail some particular
predictions of the extended CTRW-approaches.

\subsubsection{Cascaded controlling processes}

The model of cascaded control is in reality a good tool for the
analysis of the kinetic process in fractal structures. The cascade
of coupled processes can properly model the kinetic coupling of
structures of different size. The model is very useful for the
analysis of kinetics of processes in selfsimilar and complex
structures \cite{West}.

In particular, let us discuss anomalous relaxation in highly
disordered structures. In such structures the controlling processes
are often quite adequately described by the anomalous model
considered in Sec. IV.B in which the controlling evolution functions
$\hat \phi_q (\epsilon) \sim \epsilon^{\alpha_q}$ with $\alpha_q <
1$ ($q = 1,\dots ,N$) [see eq. (\ref{ctrw10})] \cite{Met}. In this
model formula (\ref{ctrw11}) predicts interesting behavior of the
total controlling evolution function $\hat \Phi (\epsilon) \sim
\epsilon^{\alpha_0},$ where $\alpha_0 = \prod_{i=1}^N \alpha_i$. In
principle, the product in equation for $\alpha_0$ is convergent for
properly behaving $\alpha_i$ as a function of $i$. Corresponding
criteria are known \cite{Math} but it is evident that the necessary
condition is $\alpha_{i \to \infty} \to 1$. This condition has
fairly clear physical interpretation: if we escribe the controlling
functions with larger numbers $i$ to the structures of smaller size,
then approaching of  $\alpha_i$ to unity results from the evident
fact that in structures of smaller size the anomalous effects,
caused by disorder of medium, are expected to be weaker. Weakness of
effects manifests itself in the reduction of processes to Markovian
with the increase of $i$, i.e. just in the relation $\alpha_{i \to
\infty} \to 1$ or $\alpha_{i} = 1$ at $i$ larger than some
characteristic number $N$.

It is also important to note another interesting prediction of the
expression (\ref{ctrw11}). It shows that in the case of cascaded
controlling process the anomaly of the processes in the cascade is
accumulated. This effect manifests itself in the decrease of
$\alpha_0$ with the increase of the number $N$ of coupled processes.
In particular, for a large number $N$ of weakly anomalous cascaded
controlling processes with $\delta_i = 1-\alpha_i \ll 1$ one gets
the value
\begin{equation}\label{dis1}
\alpha_0 \approx e^{-\sum\nolimits_{j=1}^N \delta_j} \approx e^{
-\int_0^N \!dj\, (1\!-\!\alpha_j)},
\end{equation}
which can be fairly small, corresponding to strongly anomalous
process. Formula (\ref{dis1}) can shed light on the mechanism of
formation of relaxation anomaly in disordered systems.

\subsubsection{Fluctuations of waiting time PDF matrices}

The results obtained in Sec. IV.B demonstrate that fluctuations of
waiting time PDF matrices can strongly manifest themselves in the
kinetics of CTRW-like processes.  It is, however, still worth to
point out some specific features of this effect.

1) In our consideration in Sec. IV.B these fluctuations are assumed
to result from those of the jump rate $\hat k_0$, i.e. from the
dependence $\hat k_0 ({\bf y})$. For simplicity, we have used the
model of highly localized $\hat k_0 ({\bf y}) = \hat \kappa_0 |{\bf
y}_t\rangle \langle {\bf y}_t|$. However, in accordance with
conditions of applicability of CTRW approaches discussed above in
Sec. V.A, the main conclusions on the effect of fluctuation in the
kinetics of the process remain valid for delocalized dependencies
$\hat k_0 ({\bf y})$ as well, if these dependencies are sharp enough
(for details see Sec. IV.B). In other words the obtained results are
valid for the wide class of models based on the Markovian
representation.

2) For the sake of simplicity of the analysis, the operators $\hat
\Lambda_z$ and $\hat \Lambda_y$ have been suggested to be
independent of $y$ and $z$ variables, respectively. Nevertheless,
the major part of conclusions of Sec. IV.B are valid in the case of
coupled processes in $\{y\}$- and $\{z\}$-spaces (i.e. for $[\hat
\Lambda_z,\hat \Lambda_y] \neq 0$). The most general formulation in
this case consists in replacement of the sum $\hat \Lambda_z + \hat
\Lambda_y$ by some operator $\hat \Lambda_{yz}$ in the combined
space $\{y\otimes z\}$.

3) To illustrate possible modifications of the fluctuation mechanism
within the above general formulation we will briefly discuss the
simple model which describes fluctuations occurring at moments of
jump transitions. In principle, this model corresponds to the highly
localized $z$-dependence of the operator $\hat \Lambda_y ({\bf z})$:
$\hat \Lambda_y ({\bf z})\sim \hat P_z^{tt}$, but in the end the
dependence reduces to the following modification of the operator
$\hat {\cal L}$ [see eqs. (\ref{genf4}) and (\ref{mark5})]
\begin{equation}\label{dis2}
\hat{\cal L} =  \hat k_0({\bf y})\hat P_z^{tt} \!-\!\hat P_x \hat
P_y\hat k_0({\bf y})\hat P_z^{nt}.
\end{equation}
Here $\hat P_y$ describes sudden change of ${\bf y}$-coordinate
simultaneously with the jump in $\{x\}$-space. Of course $\hat P_y$
satisfies the normalization condition $\langle {\bf y}_e |(1- \hat
P_y) = 0$, where $\langle {\bf y}_e | = \sum_i \langle {\bf y}_j |$
is the adjoint equilibrium vector in $\{y\}$-space. For example, in
the simplest variant of sudden relaxation in $\{y\}$-space $\hat P_y
= | {\bf y}_e \rangle \langle {\bf y}_e |$.

Substitution of the expression (\ref{dis2}) into the SLE
(\ref{mark2_1}) and subsequent manipulations similar to those
presented in Sec. III yield
\begin{equation}\label{dis3}
\hat {\widetilde{\mathbb G}}(\epsilon) = \langle {\bf y}_e |\hat
{\widetilde{\mathbb G}}_y(\epsilon) |{\bf y}_i\rangle,
\end{equation}
where
\begin{equation}
\hat {\widetilde{\mathbb G}}_y(\epsilon) = \hat {\widetilde{\mathbb
P}}_{\!y_{i}}\!(\epsilon) + \hat {\widetilde{\mathbb
P}}_{\!y_n}\!(\epsilon)[1-{\widetilde{\mathbb
W}}_{y_n}(\epsilon)]^{-1} \hat {\widetilde{\mathbb
W}}_{y_i}\!(\epsilon). \label{dis4}
\end{equation}
The parameters in this formula are similar to those defined in eqs.
(\ref{mark6})-(\ref{mark9}). The only difference consists in
additional term $\hat P_y$ in the expression for $\hat
{\widetilde{\mathbb W}}_{y_j}, (j = i, n),$ and in $y$-dependence of
the parameters resulting from the dependence $\hat k_0 ({\bf y})$.
According to eq. (\ref{mark8}), $\hat {\widetilde{\mathbb W}}_{y_j}
= \hat P_x\hat P_y \hat {\widetilde{w}}_{\! y_j}(\epsilon),$ where
${\widetilde{w}}_{\!y_ j} (\epsilon) = \hat k_0 (y)\langle {\bf z}_t
|\hat G_y (\epsilon)  | {\bf z}_j \rangle $ and $\hat G_y
(\epsilon)$ is obtained with $\hat k_0 ({\bf y})$. This means that
formula (\ref{dis4}) coincides with eq. (\ref{mark8}) in which $\hat
P_x$ replaced by $\hat P_{xy} = \hat P_x\hat P_y$. In other words
this formula describes jump-like migration in $\{x\otimes y\}$-space
though with jump rate $\hat k_0 (y)$ which has addition dependence
on $y$. Moreover, the expression (\ref{dis4}) can also be
represented in the form (\ref{ctrw2b}) suitable for the analysis of
the limit of diffusion-like motion governed by the operator $\hat
{\cal L}_{xy} = (1- \hat P_{xy})\hat k_0 ({\bf y})$ describing
diffusion with the coefficient $\hat D({\bf y}) \sim \hat k_0 ({\bf
y})$ in some effective potential \cite{Shu5}.

It is important to note that in this model (unlike the general model
discussed in Sec. IV.B) $\hat {\mathbb W} (t)$-fluctuations, caused
by stochastic motion in $\{y\}$-space, do not lead to the
significant and universal change of the long time behavior of
matrices $\hat w_{f_j} (t)$ defined in eq. (\ref{ctrw19}) [see eqs.
(\ref{ctrw26}) and (\ref{ctrw27})]. Some change of $\hat w_{f_j}
(t)$-behavior is, in principle, possible due to the effect of $\hat
D({\bf y}) \sim \hat k_0 ({\bf y})$-dependence, but this change is
not universal, strongly depending on specific features of $\hat k_0
({\bf y})$-behavior.

It is also worth noting that the model, which has something in
common with the considered particular variant of our general model,
is recently discussed in ref. \cite{Grig2}. Restricting ourselves to
brief comments we would only like to emphasize that, as the analysis
of above simple variant shows, the general method developed in our
work enables one to represent the results obtained in this paper in
very compact and general form.

4) From very beginning in our consideration the extensions have been
discussed within the Markovian representation reducing the problem
to the analysis of multidimensional SLE. In so doing we have not
interpreted the results in terms of the conventional approach
appealing to stochastic properties of fluctuating PDFs of
consecutive jumps $\hat {{\mathbb W}}_{\beta} (t)$ (see Sec. III).
It is clear that the case of most pronounced effect of fluctuations
corresponds to strong long time tailed correlations of the PDFs and
of course such expression can be found without difficulties. For
example, one can consider the model assuming the process of
stochastic change of $\hat {{\mathbb W}}_{\beta} (t)$ localized (in
time) near the time of fluctuation jump. In this model the
stochastic change of PDFs can be described by introducing additional
fluctuation matrix $\hat {{\mathbb W}}_{c} (t)$ with which
conventional convolution terms of type $\int_0^{t_2}\, dt_1 \hat
{{\mathbb W}} (t-t_1)\hat {{\mathbb W}} (t_1 - t_0)$ are transformed
into $\int_0^{t_2}\, dt_1 \hat {{\mathbb W}} (t-t_1)\int_0^{t_1} d
\tau \hat {{\mathbb W}}_c (t_1 - \tau) \hat {{\mathbb W}} (\tau -
t_0)$. It is clear that in the limit of very short range
dependencies $\hat {{\mathbb W}}_{c} (t)$ this model reduces to that
considered above. However, even in the most general formulation it
can be treated as a particular variant of the above-proposed
approach based on the Markovian representation.

The analysis of possible other variants of extensions which can be
analyzed within the Markovian representation will be presented
elsewhere.

\section{Conclusions}

This work concerns detailed discussion and applications of the
Markovian representation of non-Markovian CTRW-like processes and,
in particular, non-Markovian CTRW-based SLE. In reality, however the
Markovian representation, reducing the problem to the study of the
multidimensional Markovian SLE, is more general than the CTRW
approach and allows one not only to analyze the applicability of
this approach but also develop some extensions. In our work we have
considered two of them describing the effect of cascaded controlling
processes and fluctuations of jump PDFs. The number of extensions
is, however, very large. For example, interesting effects can be
predicted in variants of CTRW approaches combining two
above-mentioned extensions. Further generalizations and applications
is a subject of subsequent publications.

{\acknowledgements The work was partially supported by the Russian
Foundation for Basic Research.}


\begin{thebibliography}{99}
\bibitem{Gard} C. W. Gardiner, \,\, \textit{Handbook of Stochastic Methods}
(Springer, New York, 1985).

\bibitem{Fors} D. Forster, \textit{Hydrodynamic Fluctuations, Broken
Symmetry, and Correlation Fluctuations} (W. A. Benjamin, Inc.,
London, 1975).

\bibitem{Abr} A. \, Abragam, \,\, \textit{The principles of nuclear magnetism%
}\, (Clarendon Press, Oxford, 1961).


\bibitem{Hau} J. W. Haus and K. W. Kehr, Phys. Rep. \textbf{150}, 263 (1987).

\bibitem{Kubo} R. Kubo, J. Math. Phys. \textbf{4}, 174 (1963).

\bibitem{Bou} J.-B. Bouchaud and A. Georges, Phys. Rep. \textbf{195}, 12
(1990).

\bibitem{Met} R. Metzler and J. Klafter, Phys. Rep. \textbf{339}, 1 (2000).

\bibitem{Bark0} G. Margolin and E. Barkai, Phys. Rev. Lett. \textbf{94},
080601 (2005).

\bibitem{Bark1} G. Bel and E. Barkai, Phys. Rev. Lett. \textbf{94},
240602 (2005); Phys. Rev. \textbf{E73}, 016125 (2006).

\bibitem{Grig00} P. Allegrini, P. Grigolini, L. Palatella, and B. J. West,
Phys. Rev. \textbf{E70}, 046118 (2004).

\bibitem{Grig01} P. Allegrini, G. Aquino, P. Grigolini, L. Palatella, A. Rosa,
and B. J. West, Phys. Rev. \textbf{E71}, 066109 (2005).

\bibitem{Bark2} E. Barkai, e-print cond-mat/0608155.

\bibitem{West} B. West and W. Deering, Phys. Rep. \textbf{246}, 1
(1994).

\bibitem{Grig1} P. Grigolini, in {\it Metastability and Nonextensivity},
edited by C. Beck, G. Benedek, A. Rapisadra, and C. Tsallis (World
Scientific, Singapore, 2005).

\bibitem{Grig02} F. Barbi, M. Bologna, and P. Grigolini, Phys. Rev. Lett.
\textbf{95}, 220601 (2005).

\bibitem{Scher1} H. Scher and E. W. Montroll, Phys. Rev. \textbf{B12},
2455 (1975).

\bibitem{Wei} E. W. Montroll and G. H. Weiss, J. Math. Phys. \textbf{6},
167 (1965).

\bibitem{West1} B. J. West, M. Bologna,and P. Grigolini,
{\it Physics of Fractal Operators} (Springer-Verlag, New York,
2002).

\bibitem{Klaf1} O. Flomenborn,K. Velonia, D. Loos, S. Masuo, Mircea Cotlet,
Y. EngelBorghs, A. E. Rowan, R. J. M. Nolte, M. Van der Auweraer, F.
C. Schryver, and J. Klafter, PNAS \textbf{102}, 2368 (2005).

\bibitem{Shu1} A. I. Shushin, Phys. Rev. \textbf{E67}, 061107 (2003).

\bibitem{Sok1} A. V. Chechkin, R. Gorenflo, and I. M. Sokolov, J. Phys.
\textbf{A38}, L679 (2005).

\bibitem{Grig2} P. Allegrini, F. Barbi, P. Grigolini, and P. Paradisi,
Phys. Rev. \textbf{E73}, 046136 (2006).

\bibitem{Shu2} A. I. Shushin and V. P. Sakun, Physica \textbf{A340}, 283
(2004).

\bibitem{Shu3} A. I. Shushin, New J. Phys. \textbf{7}, 21 (2005).

\bibitem{Seki} K. Seki, M. Wojcik, and M. Tachiya, J. Chem. Phys.
\textbf{119}, 2165 (2003).

\bibitem{Bursh} A. A. Zharikov, S. I. Temkin, and A. I. Burshtein,
Chem. Phys. \textbf{103}, 1 (1986).

\bibitem{Fel} W. Feller,  \textit{An Introduction to Probability
Theory and Its Applications}\, (Wiley, New York, 1971).

\bibitem{Sok2} I. M. Sokolov, Phys. Rev. \textbf{E63}, 011104 (2000).


\bibitem{Shu4} K. Seki, A. I. Shushin, M. Wojcik, and M. Tachiya, J. Phys.
\textbf{C19}, 065117 (2007).


\bibitem{Math} I. S. Gradsteyn and I. M. Ryzhik, \textit{Tables
of Integrals, Serieses and Products}\, (Academic, San Diego, 1980).

\bibitem{Shu5} A. I. Shushin, J. Chem. Phys. \textbf{122}, 154504 (2005).

\end{thebibliography}
\end{document}